\documentclass{osa-article}
\graphicspath{{figs/}}

\journal{oe}

\articletype{Research Article}

\begin{document}

\title{Observation of spectral mode splitting in a pump-enhanced ring cavity for mid-infrared generation}

\author{Kun Huang,\authormark{1} Jiwei Gan,\authormark{1} Jing Zeng,\authormark{1} Qiang Hao,\authormark{1} Kangwen Yang,\authormark{1} Ming Yan,\authormark{2} and Heping Zeng\authormark{1,2,*}}

\address{\authormark{1}Shanghai Key Laboratory of Modern Optical System, and Engineering Research Center of Optical Instrument and System, Ministry of Education, School of Optical Electrical and Computer Engineering, University of Shanghai for Science and Technology, Shanghai 200093, China\\
\authormark{2}State Key Laboratory of Precision Spectroscopy, East China Normal University, Shanghai 200062, China}

\email{\authormark{*}hpzeng@phy.ecnu.edu.cn} 



\begin{abstract}
We report on experimental and theoretical investigation of mode-splitting dynamics in a ring cavity under the perturbation of fractional Bragg reflection from a periodically-poled nonlinear crystal. Counterintuitively, pronounced mode splitting in the spectral domain could been observed even with a tiny intensity reflection of 0.0003. The breaking of running-wave operation in the ring-cavity configuration resulted in comparable circulating fields in forward- and counter-propagation directions, which thus dramatically reduced the enhancing factor for the resonating field. In contrast, a linear cavity with intrinsically bidirectional operation was immune to the small intra-cavity reflection. Therefore, the linear-cavity layout could provide an expedient solution for a given internal reflection to obtain more stable and higher enhancement, which was confirmed by comparative studies of mid-infrared generation based on pump-enhanced difference frequency conversion. The underlying mechanism was further modeled by numerical simulations, which agreed well with experimental results. These findings could not only shed light on the understanding of the exotic feature of concatenated optical cavities, but also provide a useful guide to practical design of enhancement cavities for cavity-based frequency conversion with periodically-poled nonlinear crystals.
\end{abstract}

\section{Introduction}
Coherent laser sources in the mid-infrared spectral region provide a unique prospect for a variety of scientific, industrial and medical applications \cite{Sigrist2015, Serebryakov2010, EbrahimZadeh2008}. Specifically, the mid-infrared spectrum covers several transparent windows of the Earth's atmosphere, which facilitates the deployment of mid-infrared free-space optical communication and laser ranging, especially under bad weather conditions such as mist, fog, and haze \cite{Arnulf1957}. Additionally, the spectral region accommodates the fundamental vibrational bands (also known as `fingerprint region') for a number of constituent molecules in the gas, liquid, and solid phases, thus favoring trace gas detection and other applications of molecular spectroscopy \cite{Schliesser2012, Muraviev2018}. Currently, mid-infrared beams can be directly obtained from the laser source, such as quantum cascade lasers (QCLs) \cite{Yao2012} and fluoride fiber lasers \cite{Jackson2012, Zhu2017}. The QCLs can be tailored to operate within a broad wavelength range albeit with relatively low output power and the requirement of thermoelectrical or cryogenic cooling. The fiber lasers are favorable in terms of compactness and excellent beam quality, but the available emitted wavelengths are limited by the available gain media. 

To obtain high-power mid-infrared sources at arbitrary wavelengths, nonlinear optical frequency conversion is usually employed based on three-wave mixing \cite{Dunn1999} and super-continuum generation \cite{Yu2013, Liu2014}, amongst others. In particular, difference frequency generation (DFG) has attracted intensive attention due to its simplicity and versatility \cite{Richter2002, Galli2010, Guha2014}. Such indirect approach based on nonlinear frequency transduction has been greatly fueled by the maturation in fabricating periodically poled crystals, most notably periodically poled lithium niobate (PPLN). The involved quasi-phase-matching technique permits access to the highest nonlinear coefficient in an extended interaction length, hence significantly boosting the conversion efficiency. To go beyond the achievable mid-infrared power in the single-pass DFG, an optical cavity can be typically used to enhance ether the pump or signal/idler fields, which leads to the implementation of pump-enhanced DFG (PE DFG) \cite{Witinski2009} and optical parametric oscillators (OPOs) \cite{Bosenberg1996, Breunig2011, Sheng2012, Gu2013, Vainio2016, Smolski2016, Li2018}, respectively. 

Another desirable feature for mid-infrared laser systems is the wavelength tunability, which is essential for spectroscopic analysis. For the OPOs, since the down-converted signal is typically resonant with the optical cavity, the frequency tuning of the idler mid-infrared light has been realized by variation of cavity length, alternation of PPLN grating, or change of the crystal temperature \cite{Chang2010, Siltanen2010}. The resulting speed for wavelength scanning was thus severely limited, which in turn imposed restrictions on the response time of the spectroanalysis system. Moreover, rapid scanning is favorable to improve the detection sensitivity by suppressing the white noise through high-bandwidth data acquisition. To circumvent the limitation, PE DFG was devised, where the signal beam was provided by external injection instead of spontaneous generation within the cavity. In this case, the high-speed frequency tuning of a distributed feedback diode laser as the independent signal source can be achieved by fast modulating the operation current. Similar spirit has been explored in a cavity-enhanced DFG laser, where the signal was in resonance to the cavity rather than the pump \cite{Petrov1995}. It is worth noting that the PE DFG could also benefit the implementation of mid-infrared transmitter for free-space optical communication \cite{Buchter2009, Hao2017}, where the telecom signal with high-speed modulation could be stably down-converted into mid-infrared regime under the cavity-enhanced pumping. 

In this context, we conducted, for the first time to the best of our knowledge, a comparative study on the PE DFG for mid-infrared generation based on ring and linear cavities. Surprisingly, we found that the enhancing factor for the linear cavity was higher than that for the ring cavity due to the presence of tiny reflection induced by Bragg reflection of the PPLN crystal. The internal reflection inside the cavity would break the running-wave property of the ring cavity, thus leading to pronounced mode splitting in the spectral domain. The mode-splitting effect was manifested in the two circulating modes in the forward- and backward-propagating directions. The unusual bidirectional operation of the ring cavity due to the intrinsic reflection from the PPLN crystal was also observed in an injection-locked Ti:Sapphire laser cavity \cite{Natale2008} and in a doubly resonant OPO \cite{Rihan2011}, albeit without reporting the accompanied mode-splitting phenomenon. Therefore, it is the first time to observe the mode-splitting dynamics for a running-wave cavity. Furthermore, the underlying mechanism was theoretically investigated by using frequency-domain cavity simulation, which agreed well with the experimental observations. In the linear-cavity configuration, the PE DFG enabled to achieve nearly 60-mW mid-infrared light by injecting 1-W pump and 1-W signal beams, which was three times higher than previously reported results \cite{Witinski2009}. The counterintuitive mode-splitting dynamics in the ring cavity is fundamentally interesting. The revelation of the disparate behaviors for two types of concatenated optical cavities could give insight into the engineering of enhancement cavities for frequency conversion with periodically-poled nonlinear crystals.

\section{Experimental setup}
 
\begin{figure}[t!]
\centering
\includegraphics[width=0.85\columnwidth]{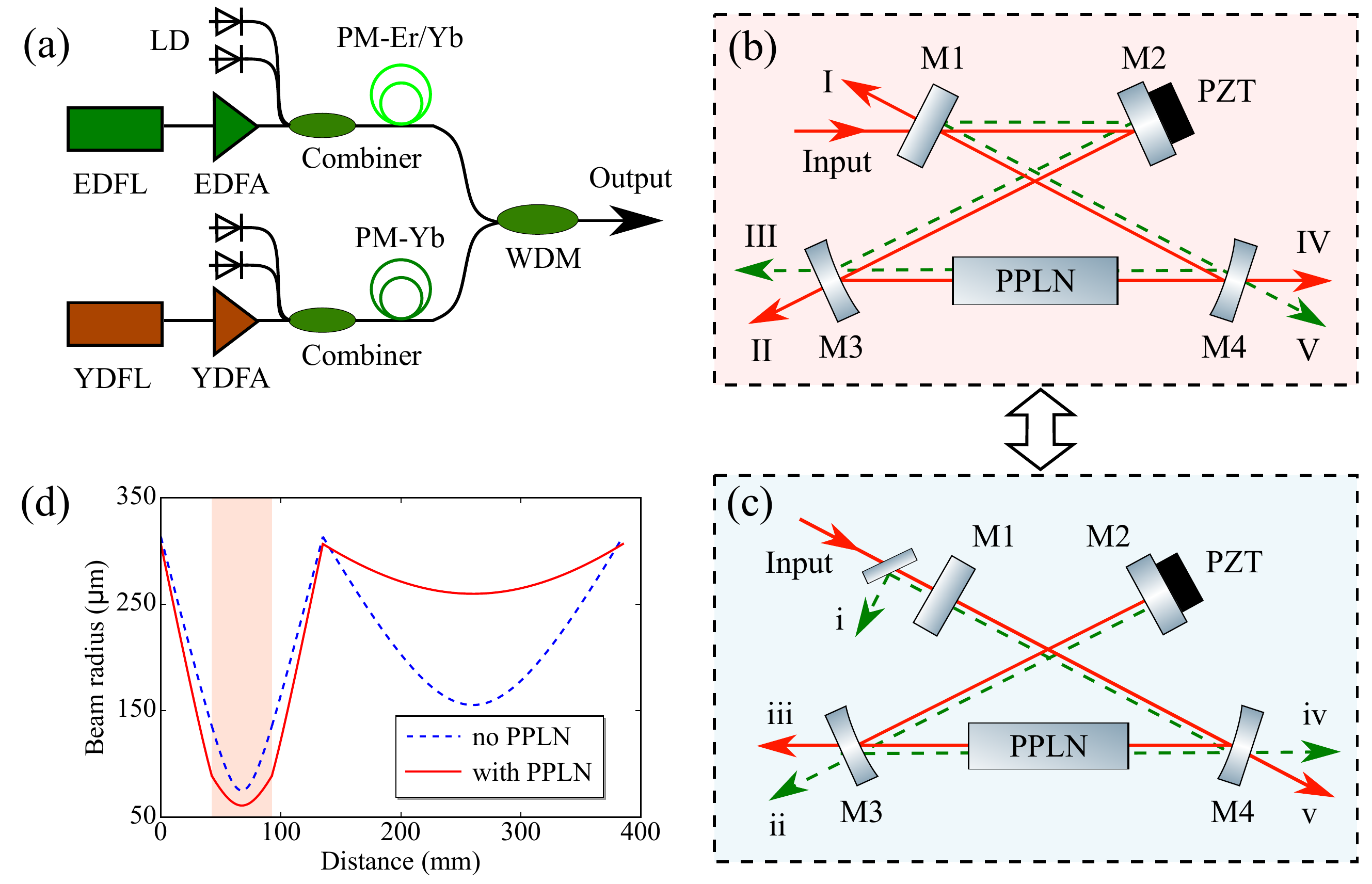}
\caption{(a) Schematic of experimental configuration for preparing pump and signal sources based on two continuous-wave single-frequency fiber lasers. Layout of a ring cavity (b) and linear cavity (c). The geometrical arrangement was engineered to keep the same beam profiles along the optical cavities. (d) Beam radius as a function of the distance from the starting point at mirror M3, for the cases with and without the presence of the nonlinear crystal inside the cavity. The shaded area indicates the occupied space by the 50-mm-length nonlinear crystal. LD: Laser diode; EDFL: Er-doped fiber laser; EDFA: Er-doped fiber amplifier; YDFL: Yb-doped fiber laser; YDFA: Yb-doped fiber amplifier; PM-Er/Yb: polarization-maintaining Er/Yb co-doped gain fiber; PM-Yb: polarization-maintaining Yb-doped gain fiber; WDM: wavelength division multiplexer; M: mirror; PZT: piezoelectric transducer; PPLN: periodically-poled lithium niobate crystal.}
\label{fig1}
\end{figure}

Figure \ref{fig1}(a) illustrates the experimental scheme for preparing pump and signal sources used to implement pump-enhanced difference frequency conversion for mid-infrared generation. The signal source was from a polarization-maintaining single frequency erbium-doped fiber laser (EDFL). A short section of 2.0-cm Er/Yb heavily doped phosphate fiber was used to achieve a large longitudinal mode spacing of the laser resonator, which enabled the single-mode operation \cite{Xu2010}. The EDFL delivered 25-mW linearly polarized light at 1550 nm. The pump source was provided by an ytterbium-doped fiber laser (YDFL). Similar to the configuration of EDFL, the YDFL consisted of a short-length Yb-doped phosphate fiber as the gain medium \cite{Xu2013}, which generated an output power of 40 mW at 1550 nm. To maintain the long-term stability of single-frequency operation, the fiber laser cavities of both EDFL and YDFL were embedded in heat sinks with a temperature fluctuation of less than 0.5 $^\circ$C. The linewidths for both fiber lasers were measured to be less than 10 kHz by using the self-heterodyne method \cite{Xu2010, Xu2013}. The frequency tuning and modulation of the signal laser could be realized by changing the currents of the pump diodes, which could facilitate fast spectral scanning in spectroscopic analysis. We note that distributed feedback diode lasers at telecommunication bands are commercially available, which can provide supreme features in frequency tuning range and modulation speed.

Then the signal and pump sources were seeded into two-stage fiber amplifier in the so-called MOPA scheme \cite{Hao2007}. Specifically, the seed laser at 1064 nm from YDFL was injected sequentially into a single-mode-fiber Yb-doped fiber amplifier (YDFA) and a double-cladding YDFA (DC-YDFA). The DC-YDFA could boost the average power to several watts. Similarly, the seed light from EDFL at 1550 nm was also amplified to be watt-scale level after a two-stage Er-doped fiber amplifier (EDFA). The amplified lasers were subsequently combined by a high-power wavelength-division multiplexer (WDM) into a common single-mode fiber. The end of the fiber was tailed with a FC/APC connector, which was followed by an achromatic aspheric lens with a 7.5-mm focal length to collimate the beam into the free space. Finally, the collimated beam was steered into an optical cavity to implement pump-enhanced nonlinear frequency conversion. The mode matching to the optical cavity was realized by optimizing the separation between the fiber end and the lens, and the distance between the collimator and the cavity. 

The cavity was comprised with four mirrors. The input coupling mirror M1 had a power transmission of 3\% at 1064 nm, and mirror M3 was coated with a high reflection of 99.9\%. Cavity mirrors M3 and M4  were calcium fluoride concave mirrors with a 100-mm radius of curvature, which had a high reflection of 99.8\% at 1064 nm and a high transmission of 95\% at 3.4 $\mu$m. The mirror M2 was attached onto a piezoelectric transducer (PZT) to implement active cavity stabilization based on the dither-locking technique. As shown in Figs. \ref{fig1}(b) and (c), the four mirrors were arranged to construct ring cavity and linear cavity to compare the pump-enhancing behaviors. Particularly, the geometrical distances between mirrors were arranged to retain the evolution of spatial modes along the optical cavity. As shown in Fig. \ref{fig1}(d), the waist radii were 61 $\mu$m and 75 $\mu$m in the presence and absence of the nonlinear crystal inside the cavity, respectively. The resulting spatial mode along the optical cavity was accommodated with the 50-mm length and 1-mm thickness for the 5\% MgO-doped PPLN crystal (HC Photonics) used in our experiment, which was highlighted with a color band in Fig. \ref{fig1}(d).

\section{Results and discussion}

\begin{figure}[t!]
\centering
\includegraphics[width=0.95\columnwidth]{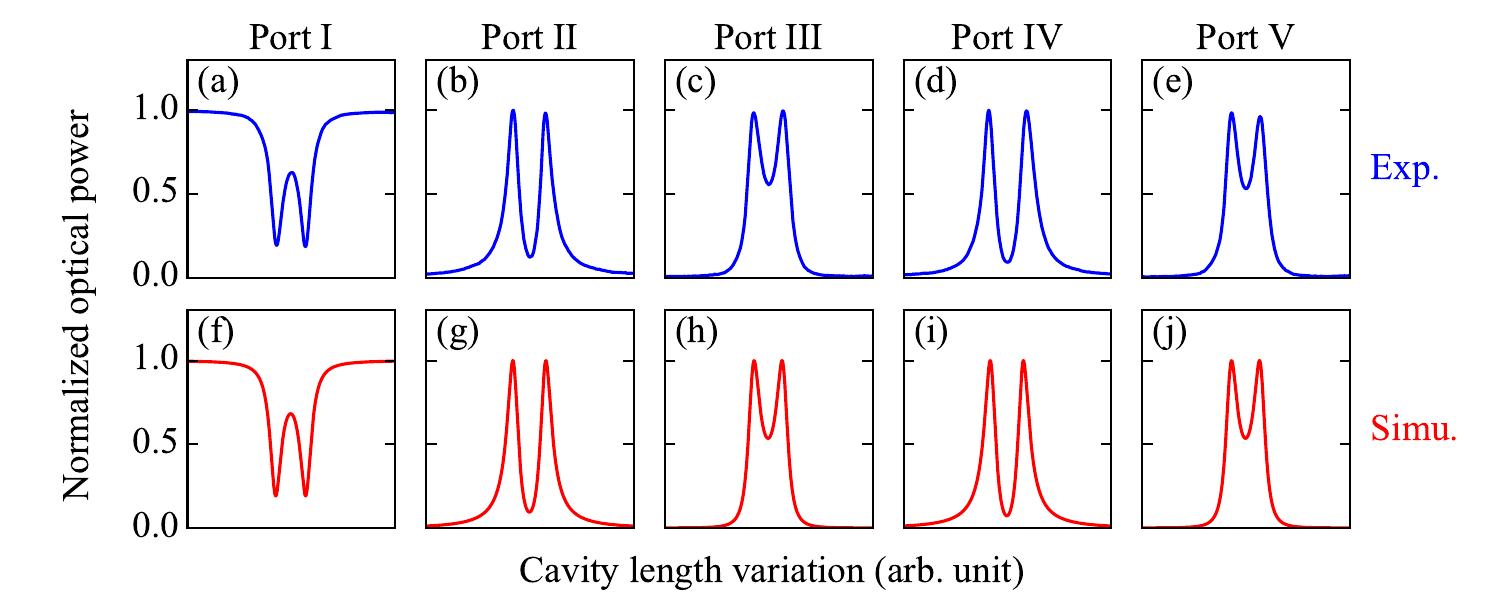}
\caption{(a-e) Experimentally recorded output signals from different ports in the ring cavity as varying the cavity length. (f-j) Numerically simulated results with an assumption of 2\% loss and 0.4\% reflection induced by the PPLN crystal.}
\label{fig2}
\end{figure}

We first investigate the ring-cavity configuration as presented in Fig. \ref{fig1}(b). Ring cavities, especially in bow-tie type, have been intensively used to realize laser resonators or optical parametric oscillators due to advantages like easy alignment, multiple access ports, and two possible circulating paths. Normally, the injected beam into the ring cavity can only propagate in one direction, for instance the forward propagation as denoted by the solid red lines in Fig. \ref{fig1}(b). Quite counterintuitively, two output beams with comparable optical power were observed at the transmitted ports, which indicated two circulating fields coexisting within the cavity. Therefore, significant amount of light would be back-reflected along the injecting path of the input signal. The resulting optical feedback necessitated the deployment of an optical isolator in the signal path to avoid perturbation on the laser amplifier and oscillator, which has also been observed in an injection-locked Ti:Sapphire laser cavity \cite{Natale2008} and in a doubly resonant OPO \cite{Rihan2011}. The unusual bidirectional operation of the ring cavity was ascribed to the Bragg reflection of the PPLN crystal. Indeed, there were no optical elements at normal incidence inside the cavity, expect for the domain walls within the grating of the PPLN crystal. It was the small index discontinuity at the boundaries of the domain walls that induced the counter-propagation scattering. Consequently, the tiny reflection at each domain wall would be coherently added under the condition of proper phase delays. The Bragg-grating behavior of the PPLN crystal was confirmed by tilting the crystal or changing the temperature. As a result, periodical change of backscattering strength could been inferred by examining power variation of the transmitted signal in two circulating directions.

To quantitatively characterize the internal refection, the reflected or transmitted beams from each cavity mirror were recorded by a free-space amplified photodetector (Thorlabs PDA36A2). As shown in Figs. \ref{fig2}(a-e), mode-splitting effect could be clearly seen in the recored spectra as varying the cavity length. Generally, the double-peak feature was more pronounced in the forward-propagating beams (at ports II and IV) than the backward-propagating ones (at ports III and V). The separation of two peaks was determined by the internal refection induced by the PPLN crystal, which could be verified by changing the orientation or temperature of the crystal. The underlying mechanism was further investigated by a theoretical cavity model that took into account of the intrinsic reflection of the PPLN crystal. Steady states of the optical cavity were numerically solved by FINESSE, a frequency-domain simulation software initially written for the design and commissioning of interferometric gravitational-wave detectors \cite{Freise2010}. Figures \ref{fig2}(f-j) present the simulation results with an assumption of 2\% loss and 0.4\% reflection due to the PPLN crystal, which agree well with the experimental observation. It is intriguing to note that even tiny fractional reflection could establish pronounced coupling between the two propagating directions, thus resulting in the mode splitting and bidirectional operation of the circulating fields.

\begin{figure}[t!]
\centering
\includegraphics[width=0.95 \columnwidth]{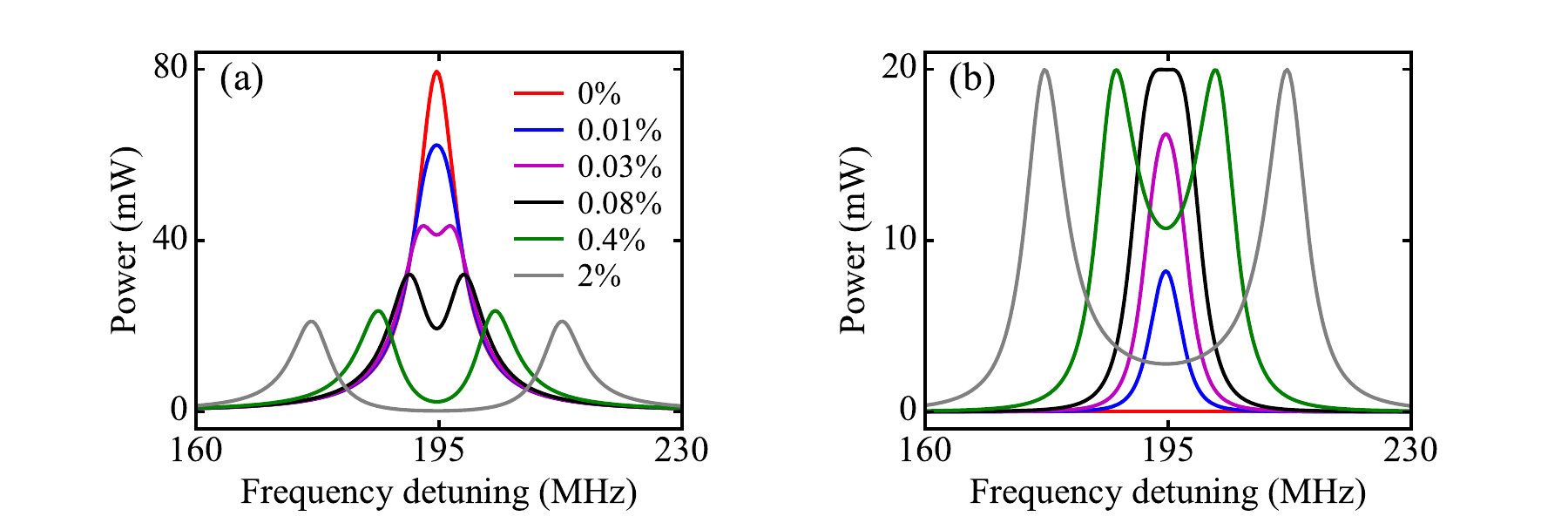}
\caption{Mode-splitting dynamics for circulating fields in the forward-propagation (a) and backward-propagation (b) directions for various internal reflection induced by the PPLN crystal. In the theoretical model, the power of the inject signal is 1 W and the intrinsic loss of the PPLN crystal is 2\%, which are comparable to experimental values.}
\label{fig3}
\end{figure}

Next we turn to investigate the mode-splitting dynamics in the presence of various amounts of the reflection. Figures \ref{fig3}(a) and (b) present the transmission spectra for circulating fields in the forward direction (\textit{e.g.} at port II) and backward direction (\textit{e.g.} at port III), respectively. In the extreme case without the internal reflection, only the forward-propagating light existed, as expected for a running-wave cavity. With a modicum of reflection as small as 0.01\%, significant back-propagating signal could be observed as shown in Fig. \ref{fig3}(b). Further augment of the reflection would lead to nearly equal power for the bidirectionally transmitted signals at the resonance condition. In the meantime, the separation between the two split peaks was enlarged. The mode-splitting effect could be observed with a small refection of 0.03\%, thus providing an agile approach to quantitatively probe the internal reflection. The sensitivity of this technique could be improved by using a cavity with a higher finesse.

\begin{figure}[t!]
\centering
\includegraphics[width=0.60 \columnwidth]{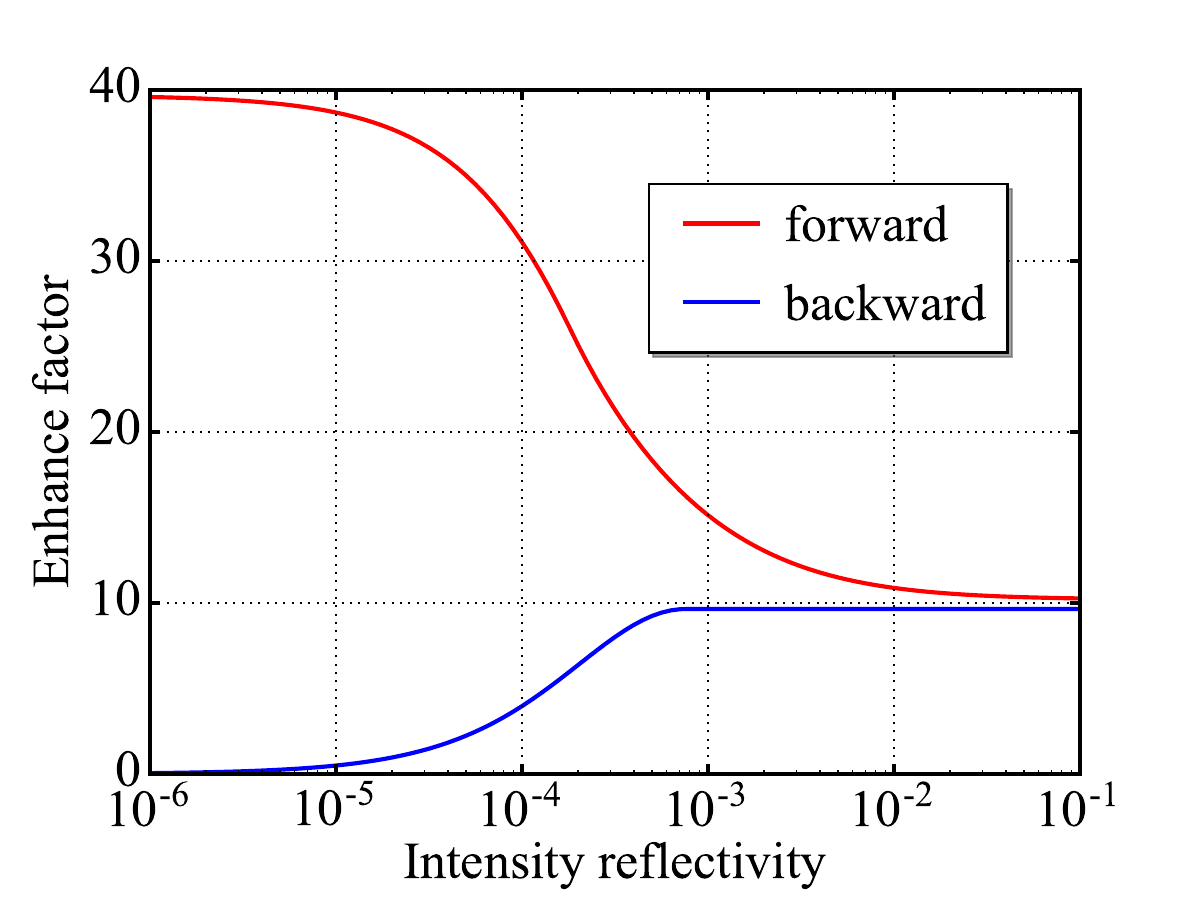}
\caption{Enhancing factor as a function of the internal intensity reflectivity induced by the PPLN crystal for the resonating optical fields in the forward and backward circulations. Significant drop of the enhancing factor can be observed even with the presence of a tiny fractional reflection.}
\label{fig4}
\end{figure}

In the following, we will study the degradation of enhancing factor $\Gamma$ with the increase of the internal reflection. For our ring cavity, the peak power enhancement can be expressed as $\Gamma = (1-R_1)/(1-\sqrt{R_1 R_2 R_3 R_4 T})^2 $, where $R_{1-4}$ are the respective intensity reflection coefficients for each mirror, and T is the intensity transmission coefficient for the intracavity element, \textit{i.e.} PPLN crystal in our case. In the absence of internal reflection, the enhancing factor was about 40 as shown in Fig. \ref{fig4}. The enhancing factor dropped rapidly with the introduction of internal reflection. Specifically, a fractional reflection of 0.01\% could result in a quarter reduction of the enhancing factor. Further increasing the reflection to 1\% would converge the bidirectional enhancement to an asymptotic value of 10. Therefore, stringent requirement of extremely low internal reflection should be fulfilled to  leverage the ring cavity to achieve a high enhancement. 

\begin{figure}[t!]
\centering
\includegraphics[width=0.95 \columnwidth]{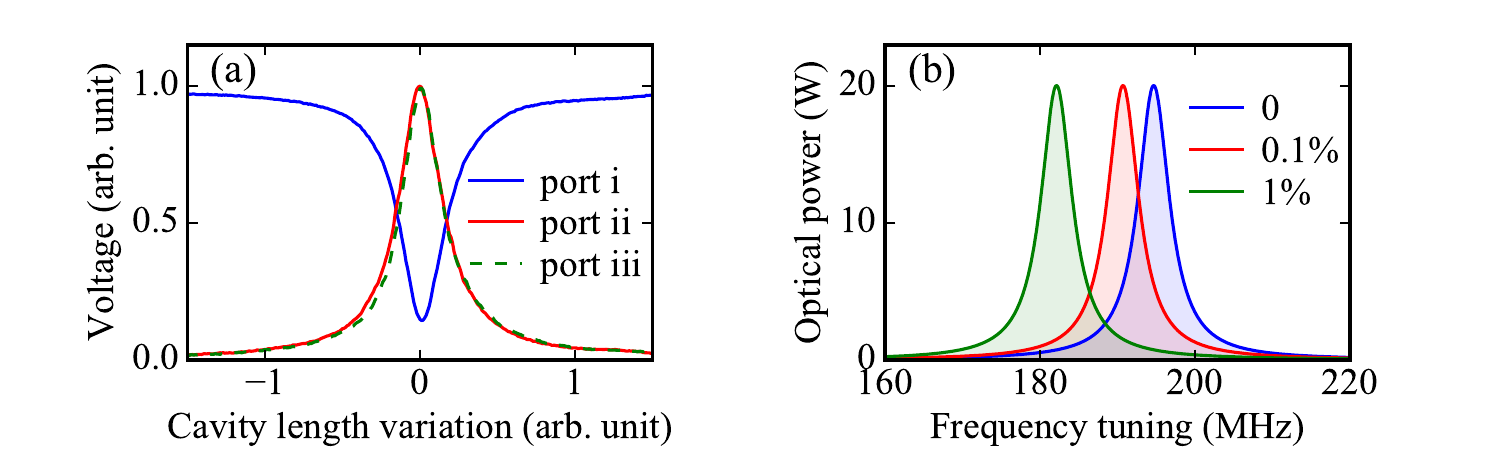}
\caption{(a) Experimentally recorded profiles in the linear-cavity configuration for the reflected signal at port i, the transmitted signal at port iii in the forward-propagating direction, and the transmitted signal at port ii in the backward-propagating direction. (b) Theoretical spectra of intra-cavity fields for various internal reflection induced by the PPLN crystal. The solid lines denote the the forward-propagating signal while the shaded areas indicate the backward-propagating signal. In the theoretical simulation, the power of the inject signal is 1 W and the intrinsic loss of the PPLN crystal is 2\%, which are close to actual values in the experiment.}
\label{fig5}
\end{figure}

For the sake of direct comparison, it is natural to adopt a linear cavity to investigate the intriguing mode-splitting effect. As depicted in Fig. \ref{fig1}(c), the geometrical dimensions of the linear cavity was designed to obtain the same resonating spatial mode as the one for the ring cavity. The intrinsically bidirectional operation for the standing-wave cavity should exclude the influence of the induced refection of the PPLN crystal inside the cavity. This expectation has been confirmed by our experimental results. Figure \ref{fig5}(a) gives the experimentally recorded signal out of the linear cavity as a function of the cavity-length variation, specifically, the reflected signal at port i, the transmitted signal at port iii in the forward-propagating direction, and the transmitted signal at port ii in the backward-propagating direction. It could be seen that no mode-splitting effect was observed in the linear-cavity configuration. Furthermore, the dependence of intra-cavity spectra on the internal reflection was investigated. As shown in Fig. \ref{fig5}(b), the solid lines denote the the forward-propagating signal while the shaded areas indicate the backward-propagating signal. We could found that the spectral profiles for the optical fields in both propagating directions were identical as expected for a linear cavity. The intra-cavity power for the bidirectionally circulating fields could be reserved at least for an internal reflection up to 1\%, thus exhibiting a sufficient tolerance to resist the Bragg reflection of the PPLN crystal. In consideration of 1-W input power, the 20-W intra-cavity peak power at the resonance indicated an enhancing factor of 20 for our linear cavity. Comparing to the ring cavity without internal reflection, the half reduction of enhancement was ascribed to the increased loss induced by double passages on the mirrors and crystal. It is worth noting that the linear cavity can still provide higher enhancing factor for an internal reflection larger than 0.04\% as shown in Fig. \ref{fig4}.

\begin{figure}[t!]
\centering
\includegraphics[width=0.7\columnwidth]{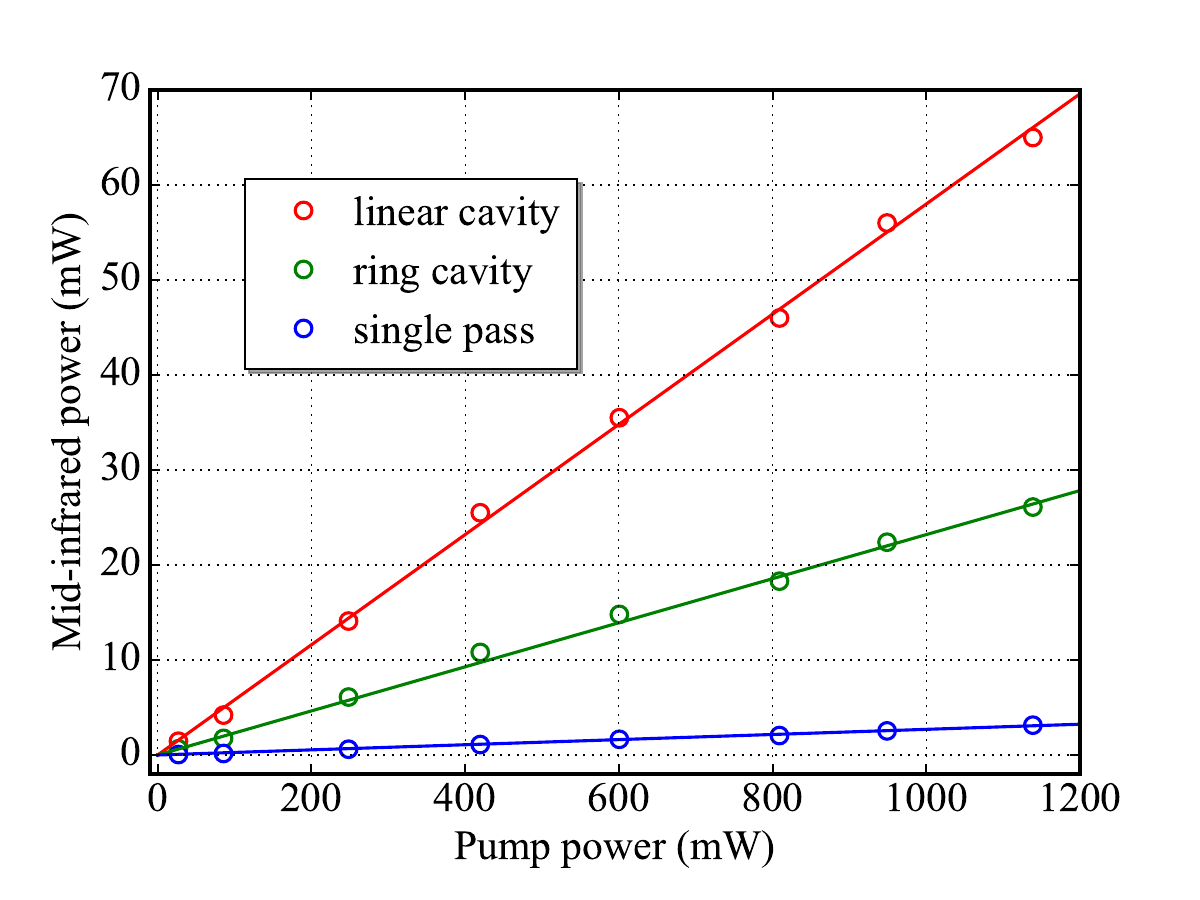}
\caption{Generated mid-infrared power as a function of pump power for single-pass, ring-cavity, and linear-cavity configurations.}
\label{fig6}
\end{figure}

Finally, the two configurations based on ring and linear cavities were used to implement pump-enhanced difference frequency conversion for mid-infrared generation. In principle, the mid-infrared power is proportional to powers of the signal at 1550 nm and the pump at 1064 nm. In the experiment, the power at 1550 nm was fixed at 1 W. In the single-pass configuration, the conversion efficiency was measured to be 0.27\% as indicated by the slope of the linear fitting in Fig. \ref{fig6}. For the ring-cavity layout, the resonator was stabilized on one peak of the split spectral modes by using the dither locking. Consequently, the cavity enhancement resulted in an almost tenfold increase of the slope efficiency to 2.32\%, as predicted by the aforementioned theoretical model. The adoption of the linear cavity enabled to reach a slope efficiency of 5.8\%, which was 20 times higher than that for the single-pass arrangement. In the case of linear cavity, the generated mid-infrared radiation was about 60 mW at the pump power of 1 W, which was three times higher than the previously reported value \cite{Witinski2009} based on a ring cavity. Therefore, for a non-negligible reflection (> 0.1\%) induced by the PPLN crystal, the use of linear cavity could provide an expedient solution to obtain more stable and higher enhancement, which could benefit the cavity-based nonlinear frequency conversion.

We note that two essential factors in the experiment allowed us to observe the unusual phenomenon. The first one was the narrow linewidth below 10 kHz for the pump laser, which was sharp enough to probe the spectrally split modes. The other one was the long 50-mm length of the crystal, which could induce higher refection due to a larger number of domain walls inside the grating. In order to minimize the reflection due to the PPLN crystal, the premiere step is to avoid normal incidence onto the crystal facets. Brewster-angle cutting and anti-reflection coating of the crystal ends could eliminate the parasitic reflection at the input surface of the crystal \cite{Natale2008}. However, the Bragg reflection could still occur inside the crystal. In this case, we could change the operation temperature of the crystal to tune the phases among the reflective fields, thus reducing the Bragg reflection. In practice, the temperature should be optimized in a trade-off with the phase-matching condition for efficiency nonlinear conversion. A more effective way to reduce the internal reflection into the cavity mode is the proper tilt of the crystal in order to realize an oblique incidence on the domain walls. This trick has been frequently used in practical experiments, which is particularly effective for a crystal with a short length. For a long crystal, the effectiveness to reduce the reflection would be compromised with the insertion loss defined by the accommodation between crystal geometric dimensions and cavity spatial modes. Another possible approach to minimize the reflection into the cavity mode might resort to the special design of oblique grating within the PPLN crystal, albeit with additional complexity of the fabrication process. 

\section{Conclusions}
In summary, we have conducted comparative investigation on the pump-enhancing effect based on ring and linear cavities, with the aim of optimizing conversion efficiency in mid-infrared generation. In particular, mode-splitting dynamics in a ring cavity was experimentally and theoretically revealed by taking into account of the small Bragg reflection induced by the PPLN crystal. Counterintuitively, a modicum of reflection as small as 0.0003 could lead to pronounced mode-splitting effect. The breaking of unidirectional operation for a running-wave cavity would result in a rapid drop of enhancing factor. In contrast, we found that the linear-cavity was immune to the perturbation of the internal reflection due to the intrinsic bidirectionality. These observations were well modeled by the numerical simulation. Moreover, performances in pump-enhanced difference frequency generation were experimentally demonstrated for the two type of cavities, which indicated that the linear cavity could enable higher conversion efficiency with the presence of non-negligible Bragg reflections (>0.1\%) of the PPLN crystal. The presented findings would help to understand the imperfections in practical enhancement cavities, which might give clues to optimize the performance for cavity-based frequency conversion with periodically-poled nonlinear crystals.

\section*{Funding.} National Key R\&D Program of China (2018YFB0407100), Program for Professor of Special Appointment (Eastern Scholar) at Shanghai Institutions of Higher Learning, Science and Technology Innovation Program of Basic Science Foundation of Shanghai (18JC1412000), National Natural Science Foundation of China (11727812).


\end{document}